\documentclass[a4paper]{aa}

\usepackage{graphicx,natbib,txfonts}

\newcommand{\Bz}{$\langle B_\mathrm{z} \rangle$}
\newcommand{\kms}{km\,s$^{-1}$}
\newcommand{\Teff}{$T_\mathrm{eff}$}
\newcommand{\vsini}{$v_\mathrm{e}\sin{i}$}

\begin{document}

   \title{Chemical spots in the absence of magnetic field in the binary \\HgMn star 66~Eridani\thanks{Based on observations collected at the European Southern Observatory, Chile (ESO program 084.D-0338).}}

   \subtitle{}

   \author{V.~Makaganiuk\inst{1}
      \and O.~Kochukhov\inst{1}
      \and N.~Piskunov\inst{1}
      \and S.~V.~Jeffers\inst{2}
      \and C.~M.~Johns-Krull\inst{3}
      \and C.~U.~Keller\inst{2}
      \and M.~Rodenhuis\inst{2}
      \and F.~Snik\inst{2}
      \and H.~C.~Stempels\inst{1}
      \and J.~A.~Valenti\inst{4}}

   \institute{Department Physics and Astronomy, Uppsala University, Box 516, 751 20 Uppsala, Sweden
   \and
Sterrekundig Instituut, Universiteit Utrecht, P.O. Box 80000, NL-3508 TA Utrecht, The Netherlands
   \and
Department of Physics and Astronomy, Rice University, 6100 Main Street, Houston, TX 77005, USA
   \and
Space Telescope Science Institute, 3700 San Martin Dr, Baltimore MD 21211, USA}

   \date{Received 00 Dec 2010 / Accepted 00 Feb 2011}

  \abstract
   {According to our current understanding, a subclass of the upper main sequence chemically peculiar stars, called mercury-manganese (HgMn), is non-magnetic. Nevertheless, chemical inhomogeneities were recently discovered on their surfaces. At the same time, no global magnetic fields stronger than 1--100~G are detected by modern studies.}
   {The goals of our study are to search for magnetic field in the HgMn binary system 66~Eri and to investigate chemical spots on the stellar surfaces of both components.}
   {Our analysis is based on high quality spectropolarimetric time-series observations obtained during 10 consecutive nights with the HARPSpol instrument at the ESO 3.6-m telescope. To increase the sensitivity of the magnetic field search we employed a least-squares deconvolution (LSD). We used spectral disentangling to measure radial velocities and study line profile variability. Chemical spot geometry was reconstructed using multi-line Doppler imaging.}
   {We report a non-detection of magnetic field in 66~Eri, with error bars 10--24~G for the longitudinal field. Circular polarization profiles also do not indicate any signatures of complex surface magnetic fields. For a simple dipolar field configuration we estimated an upper limit of the polar field strength to be 60--70~G. For the HgMn component we found variability in spectral lines of Ti, Ba, Y, and Sr with the rotational period equal to the orbital one. The surface maps of these elements reconstructed with the Doppler imaging technique, show relative underabundance on the hemisphere facing the secondary component. The contrast of chemical inhomogeneities ranges from 0.4 for Ti to 0.8 for Ba.}
  {}

   \titlerunning{Magnetic fields and variability of 66~Eri}
   \maketitle


\section{Introduction}

The phenomenon of spot formation on the stellar surfaces is known for a long time. Detailed studies of stars exhibiting surface features showed that the formation of temperature or chemical spots is closely related to the presence of magnetic field in the stellar atmospheres \citep{Donati:2009}. Nevertheless, there are exceptional cases when spectropolarimetric studies of high sensitivity detect no magnetic field but spots are still present on the stellar surface. This is observed in chemically peculiar (CP) stars of mercury-manganese (HgMn) type.

The first discovery of spots in a HgMn star was reported for $\alpha$\,And \citep{Adelman:2002}, which also shows an evolution of spot structure \citep{Kochukhov:2007}. Currently we know six other spotted HgMn stars: HR~1185 and HR~8723 \citep{Kochukhov:2005}, AR~Aur \citep{Hubrig:2006, Folsom:2010}, HD~11753, HD~53244 and HD~221507 \citep{Briquet:2010}. All these stars exhibit spots of mostly heavy chemical elements. According to theoretical models adopted for CP stars, physical processes responsible for the spot formation are associated with magnetic field. In order to enlarge the sample of spotted HgMn stars studied in detail, we have performed a comprehensive analysis of another HgMn star.

66~Eri (HR~1657, HD~32964, HIP~23794) is a well-known binary system with nearly identical components. Using the radial velocities determined by \citet{Frost:1924} and his own observations, \citet{Young:1976} found the orbital period of the system to be $P\approx5.523$~days, which was improved by the later studies \citep{Yuschenko:2001, Catanzaro:2004}. The orbital analysis yields a mass ratio $M_\mathrm{B}/M_\mathrm{A}$\footnote{Throughout our study we denote the more massive component (HgMn star, the primary) with letter ``A''. The less massive secondary component is denoted by ``B''.} very close to unity \citep{Yuschenko:2001, Catanzaro:2004}.

Photometric studies of 66~Eri are controversial. \citet{Young:1976} has estimated the times of possible eclipses based on the orbital elements of the system. His photometric observations at expected times of eclipses did not show variability. Young concluded that the orbital inclination angle is less than 80\degr. Later, \citet{Schneider:1987} detected weak variability with a period different from the orbital one. This weak photometric variability was not confirmed by \citet{Yuschenko:1999}, who in turn suggested variability with half of the orbital period.

\citet{Berghofer:1994} reported this star as an X-ray source. In the later study \citet{Hubrig:2001} detected the third, low-mass companion, which is likely to be the actual source of the X-ray radiation coming from the system. This was directly confirmed with the \textit{Chandra} X-ray imaging by \citet{Stelzer:2006}.

The early chemical analysis of 66~Eri by \citet{Young:1976} showed overabundances of Hg, Cr and Y in the primary component, while the secondary was reported to be normal. \citet{Woolf:1999} determined Hg abundance in a large number of HgMn stars, including 66~Eri. They confirmed the overabundance of mercury in the primary component. A detailed abundance analysis of 66~Eri was presented by \citet{Yuschenko:1999}, who studied chemical composition of both components of the binary system. Combining spectroscopy and photometry, they determined \Teff\,$=11100\pm100~K$ and \Teff\,$=10900\pm100~K$ for the component B and A, respectively. The similarity of the physical parameters of the components also follows from their luminosity ratio $L_\mathrm{B}/L_\mathrm{A}=0.95\pm 0.05$. The rotational velocity of both stars is equal to \vsini\,=\,17~\kms\ \citep{Yuschenko:1999}.

66~Eri is very similar to another double-line binary with an HgMn component -- AR~Aur \citep{Hohlova:1995}. Despite the fact that no magnetic field was found in AR~Aur, it shows chemical inhomogeneities \citep{Folsom:2010}. The similarity in the two HgMn SB2 stars inspired us to analyse 66~Eri with new high-precision observational data.

The paper is structured as follows. In Sect.~\ref{obs}, we describe observational material and describe the data reduction process. Section~\ref{lsd} presents the least squares deconvolution analysis. In Sect.~\ref{mf} we discuss the measurements of magnetic field in 66~Eri. Section~\ref{sd} describes the spectral disentangling. Determination of the system parameters is given in Sect.~\ref{fp}. Sect.~\ref{lpv} presents an analysis of the spectral variability. Interpretation of the line variability in terms of surface chemical maps is presented in Sect.~\ref{DI}. The paper ends with a summary of the main results and discussion in Sect.~\ref{disc}.

\section{Observations and data reduction}
 \label{obs}

Using the newly built polarimeter HARPSpol \citep{HARPSpol, Snik:2010} attached to the HARPS spectrometer \citep{HARPS} at the ESO 3.6-m telescope we have obtained high-quality observations at ten phases, covering the full orbital period of 66~Eri. The resolution of spectra is $R$\,=\,115\,000 and a typical $S/N$ is 200--300 at $\lambda\approx5200$~\AA.

The HARPS detector is a mosaic of two 2K~$\times$~4K CCDs, allowing to record 45 and 26 polarimetric echelle orders on the blue and red CCD, respectively. The calibration set included 20 bias exposures, 20 flat fields and two ThAr frames for each night. The flat fields and comparison spectra were recorded using the circular polarization mode.

All science frames of 66~Eri were obtained in circular polarization, covering the wavelength range of 3780--6913~\AA\ with a small gap at 5259--5337~\AA. Each observation of the star was split into four sub-exposures corresponding to the 45$\degr$, 135$\degr$, 225$\degr$ or 315$\degr$ position of the quarter-wave plate relative to the beam-splitter. The length of individual sub-exposures varied between 200 and 300~s, depending on the weather conditions.

The ESO reduction pipeline was not available at that time, so we employed REDUCE package \citep{reduce}, written in IDL, to perform a number of standard steps to reduce and calibrate our cross-dispersed echelle spectra. We average and subtract bias images from the master flat and science frames. To locate spectral orders, REDUCE uses a cluster analysis method in conjunction with the master flat image. For the correction of pixel-to-pixel sensitivity variations in the science images the code normalizes the master flat field. Finally, REDUCE removes the scattered light and performs the optimal extraction of the stellar spectra as described by \citet{reduce}.

HARPS spectrometer is well-known for its stability, which is of the order of 1~m\,s$^{-1}$. Taking this into account, it is sufficient for our study to use only one ThAr spectrum obtained during each night to do the wavelength calibration. Employing $\sim$\,700--900 ThAr lines we constructed a 2-D wavelength solution with an internal wavelength calibration accuracy of 18--21~m~s$^{-1}$. The final step prior to the Stokes parameters derivation is continuum normalization. The continuum level is set by dividing each spectrum by a smooth, slowly varying function, obtained by fitting the upper envelope of the blaze shape corrected, spliced spectrum.

For the derivation of Stokes~$V$ parameter we use the ratio method described by \citet{Bagnulo:2009}. This method minimizes spurious polarization effects by an appropriate combination of the two physical beams recorded for four different retarder quarter-wave plate positions. Along with the polarization we also derive a diagnostic null spectrum. It is obtained by combining the polarization signal in individual sub-exposures destructively, thus showing the residual instrumental polarization remaining after application of the ratio method. We use the null spectrum in the same analysis steps as the Stokes~$V$ spectrum, thus providing a realistic estimate of possible errors.

Combination of eight spectra in four individual sub-exposures provided a convenient possibility to detect and remove remaining cosmic hits that otherwise seriously distort the final Stokes spectra. Affected pixels are identified by their large deviation from the median and are substituted by the latter. Since our exposure times were relatively short, only 1--2 pixels required correction in each echelle order.

\section{Least-squares deconvolution}
 \label{lsd}

Previous attempts to find magnetic fields in HgMn stars set an upper limit on the longitudinal magnetic field below 100~G \citep{Shorlin:2002, Wade:2006, Auriere:2010, Folsom:2010, Makaganiuk:2011}. The fields of this strength are undetectable in individual spectral lines. In this case the sensitivity of magnetic field search can be increased by adding information from many spectral lines. Such a multi-line method has been implemented by \citet{Donati:1997} in the so-called least-squares deconvolution (LSD). This technique assumes identical shape of spectral lines and represents observations as a superposition of profiles scaled by appropriate weight. With this method it becomes possible to derive a high-precision average profile, increasing the signal-to-noise ratio by a factor of up to 1000. The latest studies of weak magnetic fields proved the LSD technique to be very effective in detecting fields weaker than 1~G \citep{Auriere:2009, Lignieres:2009}.

To construct a line mask required by LSD, we extracted the line list from VALD \citep{VALD} using an atmospheric model with the effective temperature of 10900~K and abundance of chemical elements according to the results of \citet{Yuschenko:1999} for the more massive star. We note that LSD analysis is very weakly sensitive to the choice of \Teff\ and abundances adopted for the mask.

A line mask is a set of the laboratory line wavelengths and corresponding weights. In the case of the Stokes~$I$ spectra, the weight is a residual depth of individual line, given by VALD. For the Stokes~$V$ spectra the weight is the product of central depth, the line wavelength normalized by $\lambda=4800$~\AA, and an effective Land\'e factor of a given spectral line. Spectral lines with the central depth lower than a given cutoff factor are excluded from the line mask. By setting the cutoff criterion to 0.1, we produced the line mask containing 717 spectral lines.

We use the LSD code by \citet{Kochukhov:2010} to compute LSD Stokes~$I$, $V$, and null profiles. Profiles were computed in the velocity range from $-170$ to 200~\kms\ with a step of 0.8~\kms. This value corresponds to the average pixel scale of HARPSpol spectra. The errors of LSD profiles were calculated by propagating the uncertainties of individual pixels obtained at the reduction stage.

In general there is a mismatch between a model LSD spectrum and real observations due to coarse assumptions of the LSD technique. To account for these systematic errors we scale uncertainties of the calculated LSD profiles by the reduced $\chi^2$ as described by \citet{Wade:2000}. The re-scaled LSD profile uncertainties are used for the estimation of the longitudinal magnetic field measurements.

\begin{figure}[t]
  \centering
 {\resizebox{\hsize}{!}{\rotatebox{0}{\includegraphics{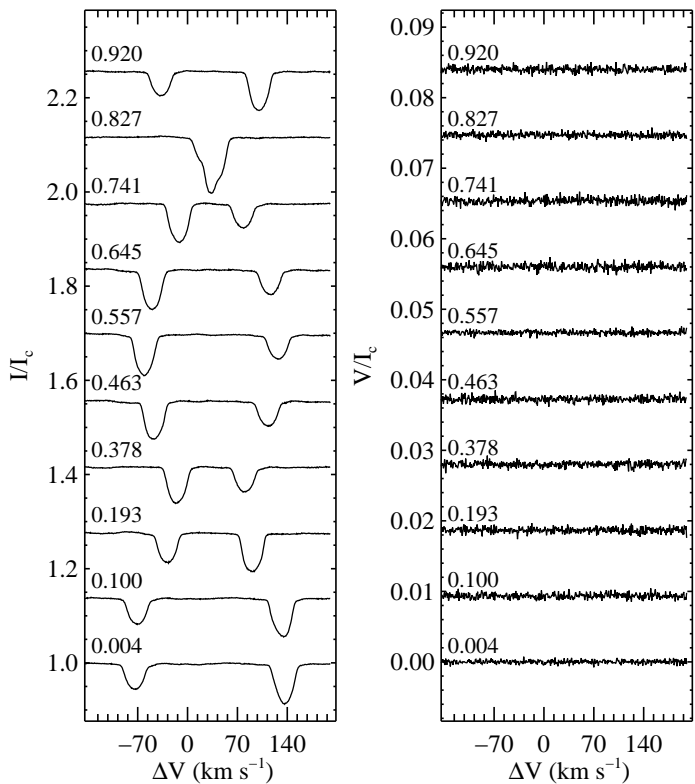}}}}
  \caption{Variation of the LSD Stokes~$I$ profiles (left panel) and LSD Stokes~$V$ (right panel) with the orbital/rotational phase. For displaying purposes we shifted the Stokes~$I$ and $V$ profiles vertically and multiplied the latter by a factor of 15. Phases are given on the left, above each profile.}
  \label{LSD}
\end{figure}

In Fig.~\ref{LSD} we present the LSD Stokes~$I$ and $V$ obtained for ten orbital phases of 66~Eri.

\section{Magnetic field measurements}
 \label{mf}

 The longitudinal magnetic field was determined from the first moment of the LSD Stokes~$V$ profile \citep{Kochukhov:2010}. Due to the contribution of blends the continuum level of raw LSD Stokes~$I$ profiles is slightly offset from unity. We use a constant factor to correct the continuum level. This factor is applied to the LSD~$V$ and null profiles. Such a correction changes the value of the longitudinal magnetic field and its error by 2--3~G.

For nine LSD~$V$ profiles we could measure the longitudinal field for each component separately. At the phase 0.827 (see Fig.~\ref{LSD}) we have measured both components as a single star. The results of the longitudinal magnetic field measurements are presented in Table~\ref{tab1}. Its columns contain the following information: the heliocentric Julian date of observation, the signal-to-noise ratio of the original spectra measured around 5200~\AA, the signal-to-noise ratio of the LSD~$V$ profiles, an orbital phase, determined from the binary solution discussed below. The next three columns represent measurements for the HgMn star (component A) as follows: the mean longitudinal magnetic field inferred from the LSD Stokes~$V$, the longitudinal field inferred from the null profile, and a False Alarm Probability (FAP, see below). The last three columns give the same information for the component B.

\begin{table*}[!t]
   \caption{Magnetic field analysis of 66 Eri.}
   \label{tab1}
   \centering
   \begin{tabular}{c l c c|r r c |r r c}
  \hline\hline
   &  &  &  & \multicolumn{3}{c}{A} & \multicolumn{3}{c}{B} \\ \cline{5-10}
HJD$-24\times 10^{5}$ & $S/N$ & $S/N$(LSD) & Phase & \Bz (V), G~ & \Bz (null), G~ & FAP$\times$10 & \Bz (V), G~ & \Bz (null), G~ & FAP$\times$10\\
  \hline
55202.725056 & 245 & 2638 & 0.193 & $	28\pm14$	& $  7\pm14$	&	4.92 & $	-26\pm18$	& $	39\pm18$	& 5.49	\\
55203.744467 & 299 & 2674 & 0.378 & $	30\pm20$	& $-22\pm20$	&	9.29 & $	16\pm17$	& $	8\pm17$		& 7.73	\\
55204.737221 & 320 & 3304 & 0.557 & $	-20\pm17$	& $ 32\pm17$	&	0.61 & $	2\pm11$		& $	-12\pm11$	& 9.63	\\
55205.755838 & 185 & 2409 & 0.741 & $	11\pm23$	& $ 61\pm23$	&	3.03 & $	-7\pm16$	& $	9\pm16$		& 2.60	\\
55206.736354 & 304 & 2671 & 0.920 & $	0\pm16$		& $ 15\pm16$	&	2.59 & $	37\pm23$	& $	27\pm23$	& 3.21	\\
55207.735135 & 315 & 2929 & 0.100 & $	-2\pm13$	& $-25\pm13$	&	0.80 & $	32\pm22$	& $	-16\pm22$	& 3.24	\\
55209.741479 & 261 & 2519 & 0.463 & $	-10\pm24$	& $  3\pm24$	&	9.75 & $	4\pm15$		& $	-13\pm15$	& 1.76	\\
55210.746427 & 216 & 2225 & 0.645 & $	-43\pm26$	& $  0\pm26$	&	2.11 & $	-19\pm17$	& $	5\pm17$		& 9.74	\\
55211.750571 & 241 & 2965 & 0.827 & $	8\pm12$		& $  0\pm12$	&	3.62 & $	8\pm12$		& $	0\pm12$		& 3.62	\\
55212.727307 & 380 & 3901 & 0.004 & $	-1\pm14$	& $ 40\pm14$	&	8.28 & $	1\pm11$		& $	-14\pm11$	& 6.58	\\
  \hline
   \end{tabular}
\end{table*}

The errors of the magnetic field measurements were obtained by the standard error propagation, using the uncertainties of the LSD~$V$ profiles. The mean error of longitudinal field measurements is $\sim$14~G for the HgMn component and $\sim$20~G for the secondary. For the HgMn star the longitudinal field value reaches the confidence level of 2$\sigma$ only at the orbital phase 0.193. For the secondary star the measurements of magnetic field do not exceed $\sim$1.5$\sigma$ level for most of the orbital phases. Thus, none of the individual \Bz\ measurements indicate the presence of magnetic field in 66~Eri.

\begin{figure}[!t]
  \centering
 {\resizebox{\hsize}{!}{\rotatebox{90}{\includegraphics{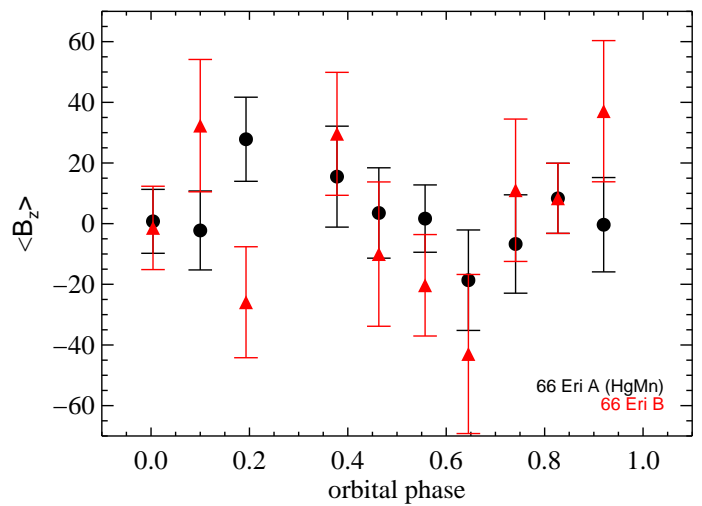}}}}
  \caption{Measurements of the longitudinal magnetic field with the corresponding error bars. Circles show the \Bz\ values for the primary component (HgMn star) and triangles represent the measurements for the secondary.}
  \label{Bz_phase}
\end{figure}

Our magnetic field measurements for both components with their respective error bars are presented in Fig.~\ref{Bz_phase} as a function of orbital phase. The actual values of the longitudinal field measurements do not exceed the level of 3$\sigma$, so the periodic-like fluctuation of magnetic field measurements in both components are fully attributed to errors.

Assuming a dipolar configuration of the magnetic field we made an assessment of the upper limit of its strength, which would be consistent with the entire set of \Bz\ measurements. Using a relation between the longitudinal field and parameters of an oblique dipole \citep{Leroy:1994}, we estimated the strength of dipolar magnetic field to be 60--70~G with an uncertainty of 50--70~G for both stars. 
However, based on the reduced $\chi^{2}$ statistics, the dipolar fit has no advantage over the null hypothesis.

At the same time, much more complex magnetic fields can exist in the atmospheres of both stars. To investigate possible presence of such complex magnetic structures, which might yield negligible \Bz, we employed the FAP analysis of LSD profiles. FAP is a $\chi^2$ statistical estimate, which assesses if the observed structure in the Stokes~$V$ line profile is produced by a random noise. Following the definition by \citet{Donati:1997}, these numbers should be interpreted as follows. \textit{No detection} if FAP~$\,>\,$10$^{-3}$, \textit{marginal detection} if 10$^{-3}\le\,$FAP$\le\,$10$^{-5}$ and \textit{definite detection} if FAP$\,<\,$10$^{-5}$. The smallest FAP obtained for HgMn star is 0.08. Thus, none of the measurements indicate the presence of a magnetic signal. A visual examination of each LSD Stokes~$V$ profile also do not suggest a magnetic field detection. 

To make sure that the LSD circular polarization profiles were not affected by a spurious polarization, we analysed LSD null spectra at each phase. The \Bz\ and FAP measurements were done in the same way as for LSD Stokes~$V$ profiles. The \Bz\ errors for the null spectra are fully consistent with those inferred from the circular polarization profiles. The absence of signal in the null spectrum means that our scientific data is not visibly affected by significant spurious polarization.

\section{Spectral disentangling}
 \label{sd}

Due to the orbital motion, the relative shift of the spectral lines of the components in the composite spectrum varies from zero to $\sim$200~\kms, which can be seen from the LSD~$I$ profiles in Fig.~\ref{LSD}. In addition to this, the primary component shows an intrinsic line profile variability. Since one of the main aims of this study is to investigate line profile variability in 66~Eri, it is important to separate the effect of the spectral line blending due to the orbital motion from an intrinsic variability.

\begin{figure*}[t]
  \centering
 {\resizebox{\hsize}{!}{\rotatebox{90}{\includegraphics{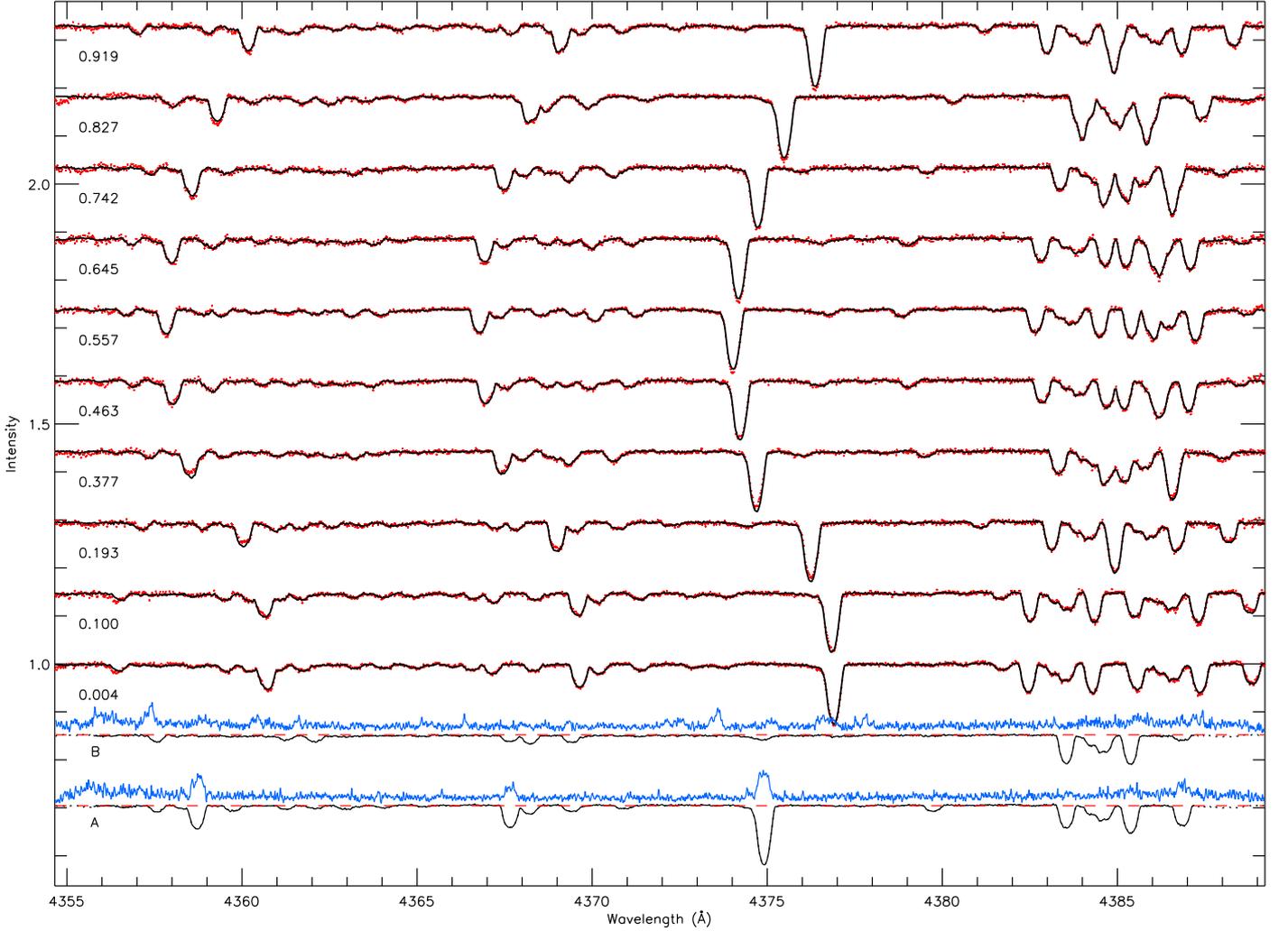}}}}
  \caption{Spectral disentangling procedure applied to 66~Eri. Dots show the observations and the solid line is the model spectrum for corresponding phase. Black solid lines at the bottom represent disentangled spectra for the primary and secondary components. The lines above the continuum level represent the standard deviations in the reference frames of components A and B. Peaks at the location of spectral lines in the standard deviation indicate the presence of intrinsic variability in a given line (\ion{Y}{ii}~4358.728~\AA, \ion{Ti}{ii}~4367.652~\AA, and \ion{Y}{ii}~4374.935~\AA\ in this spectral interval).}
  \label{sdplot}
\end{figure*}

To separate different variability effects we employ the procedure of spectral disentangling described by \citet{Folsom:2010}. It combines information from all orbital phases and yields radial velocities (RV) for both components, their average separated spectra and standard deviation in the reference frame of each component. Fig.~\ref{sdplot} shows an example of spectral disentangling in the region 4355--4380~\AA\ containing several variable lines.

\begin{figure}[!t]
  \centering
 {\resizebox{\hsize}{!}{\rotatebox{0}{\includegraphics{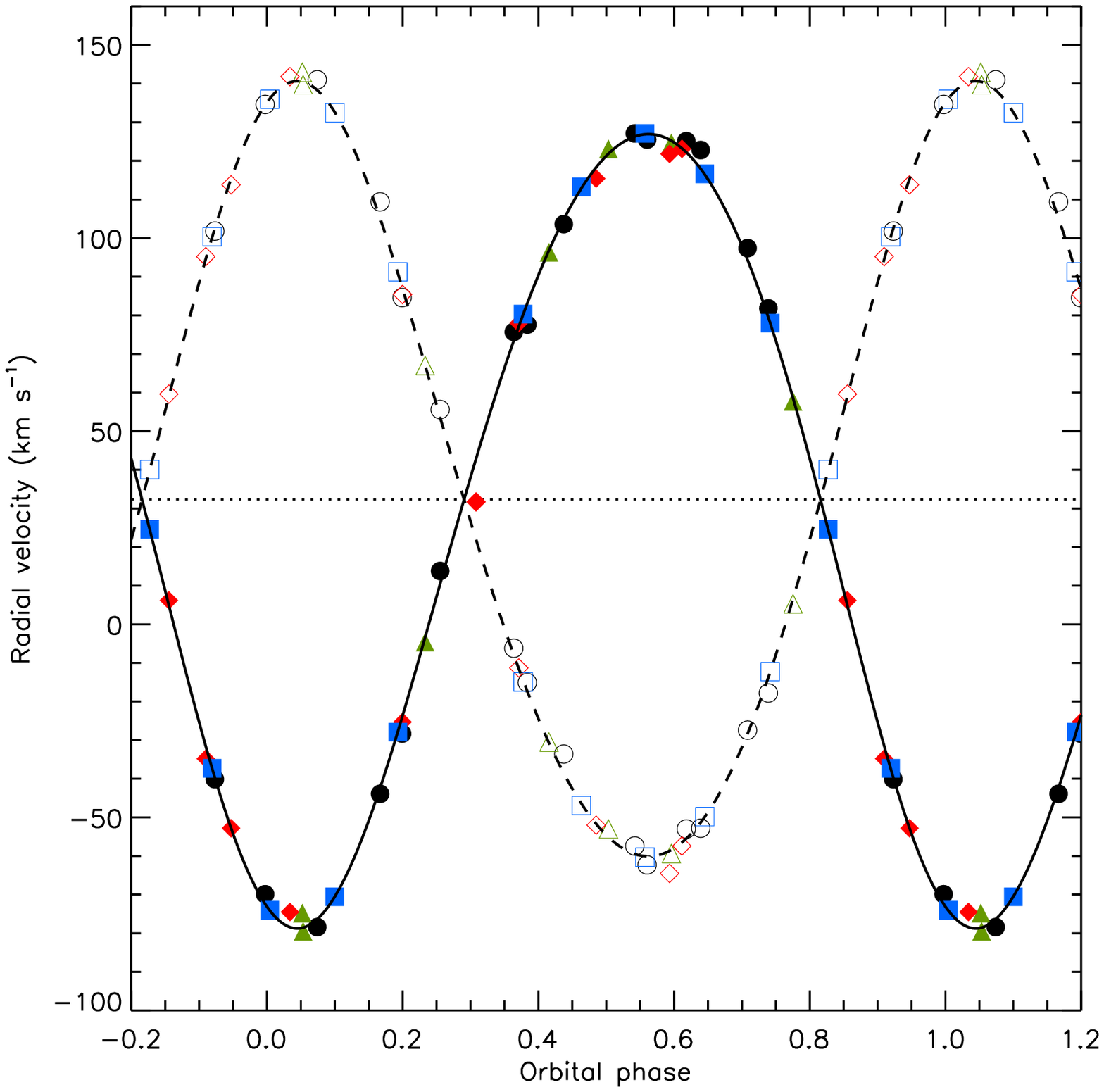}}}}
  \caption{Comparison of the radial velocities from different studies with our orbital solution. Filled symbols correspond to the secondary component and open ones to the primary. Circles represent measurements from \citet{Young:1976}, triangles represent velocities from \citet{Yuschenko:1999, Yuschenko:2001}, diamonds are from \citet{Catanzaro:2004}, and squares represent our measurements. The lines show theoretical radial velocity curve for the secondary (solid) and primary (dashed).}
  \label{orvs}
\end{figure}

The spectral disentangling procedure provided a new set of high-precision RV measurements for both components of 66~Eri. These data, presented in Table~\ref{tab2}, have typical error bars of 0.1--0.2~\kms. We combined our RVs with the velocities published by \citet{Young:1976}, \citet{Yuschenko:1999, Yuschenko:2001}, and \citet{Catanzaro:2004}, deriving improved orbital parameters of the system. The new orbital elements are listed in Table~\ref{tab3}, while Fig.~\ref{orvs} compares the respective predicted RV curves with the actual measurements. The standard deviation of the orbital fit is 2.5~\kms\ for the primary and 2.8~\kms\ for the secondary. Compared to the previous orbital analysis by \citet{Catanzaro:2004}, we significantly improved the accuracy of the orbital period and of the velocity amplitudes, obtaining a mass ratio with the precision better than 1\%.

\begin{table}[t]
   \caption{Radial velocities of the primary and secondary components of 66 Eri.}
   \label{tab2}
   \centering
   \begin{tabular}{c c r r}
  \hline\hline
   &  & \multicolumn{1}{c}{A} & \multicolumn{1}{c}{B} \\ \cline{3-4}
HJD$-24\times 10^{5}$ & Phase & $V_\mathrm{r}$, \kms\ & $V_\mathrm{r}$, \kms\\
  \hline
55202.725056 & 0.193 & $ 91.26\pm0.19$	& $-27.92\pm0.18$	\\
55203.744467 & 0.378 & $-14.91\pm0.20$	& $ 80.34\pm0.14$	\\
55204.737221 & 0.557 & $-60.30\pm0.14$	& $127.11\pm0.17$	\\
55205.755838 & 0.741 & $-12.25\pm0.13$	& $ 77.98\pm0.16$	\\
55206.736354 & 0.920 & $100.36\pm0.11$	& $-37.29\pm0.11$	\\
55207.735135 & 0.100 & $132.44\pm0.15$	& $-70.55\pm0.13$	\\
55209.741479 & 0.463 & $-46.91\pm0.21$	& $113.26\pm0.25$	\\
55210.746427 & 0.645 & $-49.79\pm0.09$	& $116.64\pm0.16$	\\
55211.750571 & 0.827 & $ 40.04\pm0.09$	& $ 24.58\pm0.21$	\\
55212.727307 & 0.004 & $-74.02\pm0.08$	& $135.94\pm0.08$	\\
  \hline
   \end{tabular}
\end{table}

\begin{table}[t]
   \caption{Results of the spectroscopic orbital analysis of 66~Eri.}
   \label{tab3}
   \centering
   \begin{tabular}{l r}
  \hline\hline
Parameter & Value~~~~~~~ \\
  \hline
$P_\mathrm{orb}$ (d)									& 5.5226013$\pm$0.0000020	\\
$T_\mathrm{0}$ (d)									& 41356.499$\pm$0.017		\\
$K_\mathrm{A}$ (\kms)								& 102.83 $\pm$0.20			\\
$K_\mathrm{B}$ (\kms)								& 100.35 $\pm$0.19			\\
$\gamma$ (\kms)										& 32.28 $\pm$0.10			\\
$e$													& 0.0844 $\pm$0.0013		\\
$\omega$ (\degr)										& 160.9  $\pm$1.1			\\
$M_\mathrm{B}/M_\mathrm{A}$						& 0.976$\pm$0.003			\\
$M_\mathrm{B}/M_\mathrm{\sun}\sin^3{i_\mathrm{orb}}$	& 2.345$\pm$0.008			\\
$M_\mathrm{A}/M_\mathrm{\sun}\sin^3{i_\mathrm{orb}}$	& 2.403$\pm$0.008			\\
$a_\mathrm{B}/R_\mathrm{\sun}\sin{i_\mathrm{orb}}$	& 11.185$\pm$0.022			\\
$a_\mathrm{A}/R_\mathrm{\sun}\sin{i_\mathrm{orb}}$	& 10.916$\pm$0.020			\\
  \hline
   \end{tabular}
\end{table}

\section{Fundamental parameters of the components}
 \label{fp}

The very similar properties of the 66~Eri components make it challenging to determine their parameters individually. The previous detailed study of the system by \citet{Yuschenko:1999} suggested that the slightly more massive component, the HgMn star, is also more luminous, but cooler by $200\pm140$~K than the other star. The sign of this spectroscopically determined temperature difference is impossible to reconcile with the main sequence status of co-evolving binary components.

\begin{figure}[!t]
  \centering
 {\resizebox{\hsize}{!}{\rotatebox{0}{\includegraphics{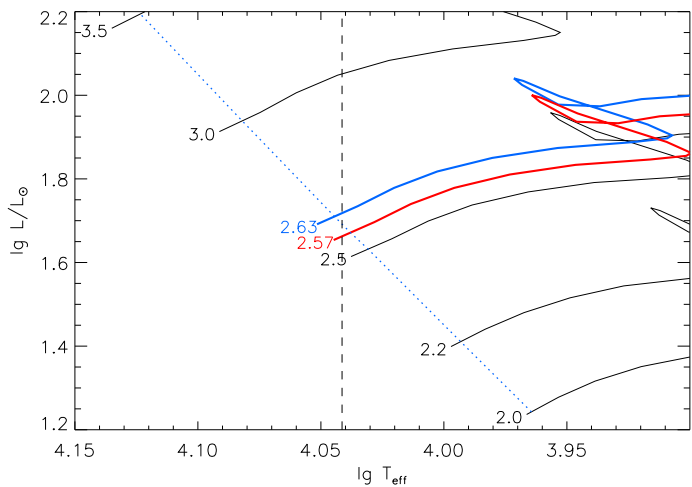}}}}
  \caption{Luminosity-effective temperature diagram constructed for the components of 66~Eri with the evolutionary calculations by \citet{Girardi:2000}. The thin solid curves represent the original evolutionary tracks, while thick curves show the tracks interpolated for the final component masses. The dotted line shows the isochrone for $\log t$\,=\,7.47 yr. The vertical dashed line corresponds to mean effective temperature of the components, $T_{\rm eff}\,=\,11000~K.$}
  \label{fig:params}
\end{figure}

We have redetermined the fundamental parameters of 66~Eri using Padova evolutionary tracks \citep{Girardi:2000} and taking advantage of the improved orbital parameters determined in our study and an accurate Hipparcos parallax, $\pi$\,=\,$10.56\pm0.34$ mas, available for the system \citep{Van-Leeuwen:2005}. The primary component mass and the age of the system were found by interpolating in the set of evolutionary tracks for $Z=0.018$ in order to match the total Hipparcos luminosity, $L/L_{\sun}$\,=\,$98.2\pm6.3$, and the mean effective temperature, $\langle T_{\rm eff} \rangle$\,=\,11000~K, determined by \citet{Yuschenko:1999} from the photometric properties and spectral energy distribution of 66~Eri. The outcome of this analysis is illustrated in the upper panel of Fig.~\ref{fig:params}. Table~\ref{tbl:params} summarizes the resulting fundamental parameters of the components, with error bars reflecting the uncertainty of the total luminosity. Effective temperatures of the components are given without error bars because they are mainly sensitive to $\langle T_{\rm eff} \rangle$, for which no uncertainty was provided by \citet{Yuschenko:1999}.

Our analysis of the H-R diagram position of the components of 66~Eri suggests that the primary is more massive, hotter, and more luminous ($L_{A}/L_{B}$\,=\,1.093), as one would expect for two co-evolving main sequence stars. Our temperature difference, $T_{\rm eff}(A)-T_{\rm eff}(B)=163$~K, is opposite to what was suggested by \citet{Yuschenko:1999}. Nevertheless, the results of their abundance analysis of 66~Eri components are unlikely to change substantially due to $\sim$\,200~K change of $T_{\rm eff}$.

The evolutionary masses and radii allow us to determine the orbital and rotational inclination angles, testing synchronization of the system. We found the projected rotational velocities, $v_{\rm e}\sin i_{\rm rot}$\,=\,$17.1\pm0.2$ and $16.9\pm0.2$ km\,s$^{-1}$ for the component A and B, respectively, from fitting the profiles of $\sim$\,20 unblended lines in the 4400--4600~\AA\ wavelength region. These results are in excellent agreement with $v_{\rm e}\sin i_{\rm rot}$\,=\,17 km\,s$^{-1}$ given by \citet{Yuschenko:1999} for both stars. Since the radii are known, the oblique rotator relation yields $i_{\rm rot}$\,$\sim$\,74\degr\ assuming $P_{\rm rot}=P_{\rm orb}=5\fd5226$ (see lower panel in Fig.~\ref{fig:params}). On the other hand, from the orbital solution and evolutionary masses $i_{\rm orb}=76\degr$. This convergence to nearly identical inclination angles derived using independent methods is a strong indication that the rotation of the 66~Eri components and their orbital motion are coaxial and synchronized.

\begin{table}[!t]
   \caption{Fundamental parameters of the primary and secondary components of 66~Eri.}
   \label{tbl:params}
   \centering
   \begin{tabular}{lcc}
  \hline\hline
Parameter & A & B \\
  \hline
$\log t$ (yr)				& \multicolumn{2}{c}{$7.47\pm0.23$}		\\
$M/M_{\sun}$					& $2.629\pm0.032$	& $2.566\pm0.032$	\\
$R/R_{\sun}$					& $1.948\pm0.063$	& $1.919\pm0.061$	\\
$L/L_{\sun}$					& $51.3\pm3.3$		& $46.9\pm3.0$		\\
$T_{\rm eff}$ (K)			& 11077				& 10914				\\
$\log g$						& $4.28\pm0.03$		& $4.28\pm0.03$		\\
$v_{\rm e}\sin i\ $ (km s$^{-1}$)& $17.1\pm0.2$	& $16.9\pm0.2$		\\
  \hline
   \end{tabular}
\end{table}

\section{Line profile variability}
 \label{lpv}

The line profile variability search was performed based on the disentangled spectra of both components. We carefully examined each spectral region, excluding hydrogen lines and the regions populated by the telluric lines. To diagnose the variability in spectral lines we computed standard deviation in the reference frame of each component. It is shown in Fig.~\ref{sdplot} by a line above the average spectra of components A and B. An excess of standard deviation at the positions of spectral lines indicate their variability.

The variability search in spectral lines was performed for both components. We confirm that the secondary component does not exhibit any variability, while for the primary star we found variability in the spectral lines of \ion{Ti}{ii}, \ion{Sr}{ii}, \ion{Y}{ii}, and \ion{Ba}{ii}. This variability is generally weaker than found in other spotted HgMn stars. We suspect that a weak variability is also present in some Cr lines.

Earlier studies of HgMn stars \citep{Adelman:2002, Kochukhov:2005} focused on the variability of \ion{Hg}{ii} line at $\lambda$ 3984~\AA. Our analysis of this spectral line shows that, despite being rather strong, it shows extremely weak variability in 66~Eri A.

The full list of variable lines is given in Table~\ref{tab:varprofs}. The first column of this table gives a line identification, the second lists the corresponding laboratory wavelength. The last two columns provide the same information for occasional blending lines. Lines with marginal variability are marked with italics.

\begin{table}
   \caption{The list of variable lines in the primary component of 66~Eri. Bold font marks spectral lines used for Doppler imaging and italics indicate lines with marginal variability.}
   \label{tab:varprofs}
   \centering
   \begin{tabular}{l c l c}
  \hline\hline
     Ion     & $\lambda$ (\AA)   	&      Ion     & $\lambda$ (\AA) \\
  \hline
\ion{Hg}{ii} & \textit{3983.931}	&				&			\\
\ion{Ti}{ii} & 4028.338			& \ion{Si}{ii} 	& 4028.465	\\
\ion{Ti}{ii} & 4053.821			& \ion{Cr}{ii} 	& 4054.076	\\
\ion{Sr}{ii} & \bf{4077.709}		&				&		  	\\
\ion{Y}{ii}  & 4124.907		 	& \ion{Fe}{ii} 	& 4124.787	\\
\ion{Cr}{ii} & \textit{4275.567}	&				&			\\
\ion{Ti}{ii} & \bf{4163.644}		&				&			\\
\ion{Ti}{ii} & 4171.904		 	&			 	&			\\
\ion{Y}{ii}  & 4177.529		 	& \ion{Fe}{ii}	& 4177.692	\\
\ion{Y}{ii}  & 4199.277		 	& \ion{Fe}{i}	& 4199.095	\\
\ion{Y}{ii}  & \bf{4204.695}  	&			 	&			\\
\ion{Sr}{ii} & \bf{4215.519}  	&			 	&			\\
\ion{Y}{ii}  & 4235.729		 	& \ion{Fe}{i}	& 4235.937	\\
\ion{Ti}{ii} & \bf{4290.215}  	&			 	&			\\
\ion{Y}{ii}  & 4309.631		 	&			 	&			\\
\ion{Y}{ii}  & 4358.728		 	&			 	&			\\
\ion{Ti}{ii} & 4367.652		 	&			 	&			\\
\ion{Y}{ii}  & 4374.935		 	& \ion{Ti}{ii}	& 4374.816	\\
\ion{Ti}{ii} & 4386.846		 	&			 	&			\\
\ion{Y}{ii}  & 4398.013		 	& \ion{Ti}{ii}	& 4398.292	\\
\ion{Ti}{ii} & \bf{4421.938}  	&			 	&			\\
\ion{Y}{ii}  & 4422.591		 	&			 	&			\\
\ion{Ba}{ii} & \bf{4554.029}		&			 	&			\\
\ion{Ti}{ii} & \bf{4563.757}		&			 	&			\\
\ion{Ti}{ii} & \bf{4571.971}	 	&			 	&			\\
\ion{Ti}{ii} & 4589.947		 	& \ion{Cr}{ii}	& 4589.901	\\
\ion{Cr}{ii} & \textit{4592.049}	&				&			\\
\ion{Cr}{ii} & \textit{4634.070}	&				&			\\
\ion{Y}{ii}  & 4682.324			&			 	&			\\
\ion{Ti}{ii} & \bf{4779.985}  	&			 	&			\\
\ion{Ti}{ii} & 4805.085		 	&			 	&			\\
\ion{Cr}{ii} & \textit{4812.337}	&				&			\\
\ion{Y}{ii}  & 4823.304		 	&				&			\\
\ion{Y}{ii}  & 4883.684		 	&			 	&			\\
\ion{Y}{ii}  & 4900.120		 	& \ion{Ba}{ii}	& 4899.929	\\
\ion{Ti}{ii} & 4911.195		 	&			 	&			\\
\ion{Ba}{ii} & \bf{4934.076} 		&			 	&			\\
\ion{Y}{ii}  & 4982.129		 	&			 	&			\\
\ion{Y}{ii}  & \bf{5087.416}  	&			 	&			\\
\ion{Y}{ii}  & 5119.112		 	&			 	&			\\
\ion{Y}{ii}  & 5123.211		 	&			 	&			\\
\ion{Ti}{ii} & \bf{5129.156}  	&			 	&			\\
\ion{Ti}{ii} & \bf{5185.902}  	&			 	&			\\
\ion{Ti}{ii} & \bf{5188.687}  	&			 	&			\\
\ion{Y}{ii}  & \bf{5196.422}  	&			 	&			\\
\ion{Y}{ii}  & \bf{5200.406}  	&			 	&			\\
\ion{Y}{ii}  & \bf{5205.724}  	&			 	&			\\
\ion{Y}{ii}  & 5473.388		 	&			 	&			\\
\ion{Y}{ii}  & 5480.732		 	&			 	&			\\
\ion{Y}{ii}  & \bf{5497.408}  	&			 	&			\\
\ion{Y}{ii}  & 5509.895		 	&			 	&			\\
\ion{Y}{ii}  & 5521.555		 	&			 	&			\\
\ion{Y}{ii}  & 5544.611		 	&			 	&			\\
\ion{Y}{ii}  & 5546.009		 	&			 	&			\\
\ion{Y}{ii}  & 5662.925		 	&			 	&			\\
\ion{Y}{ii}  & 6795.414		 	&			 	&			\\
  \hline
   \end{tabular}
\end{table}

\begin{figure*}[t]
  \centering
 {\resizebox{\hsize}{!}{\rotatebox{90}{\includegraphics{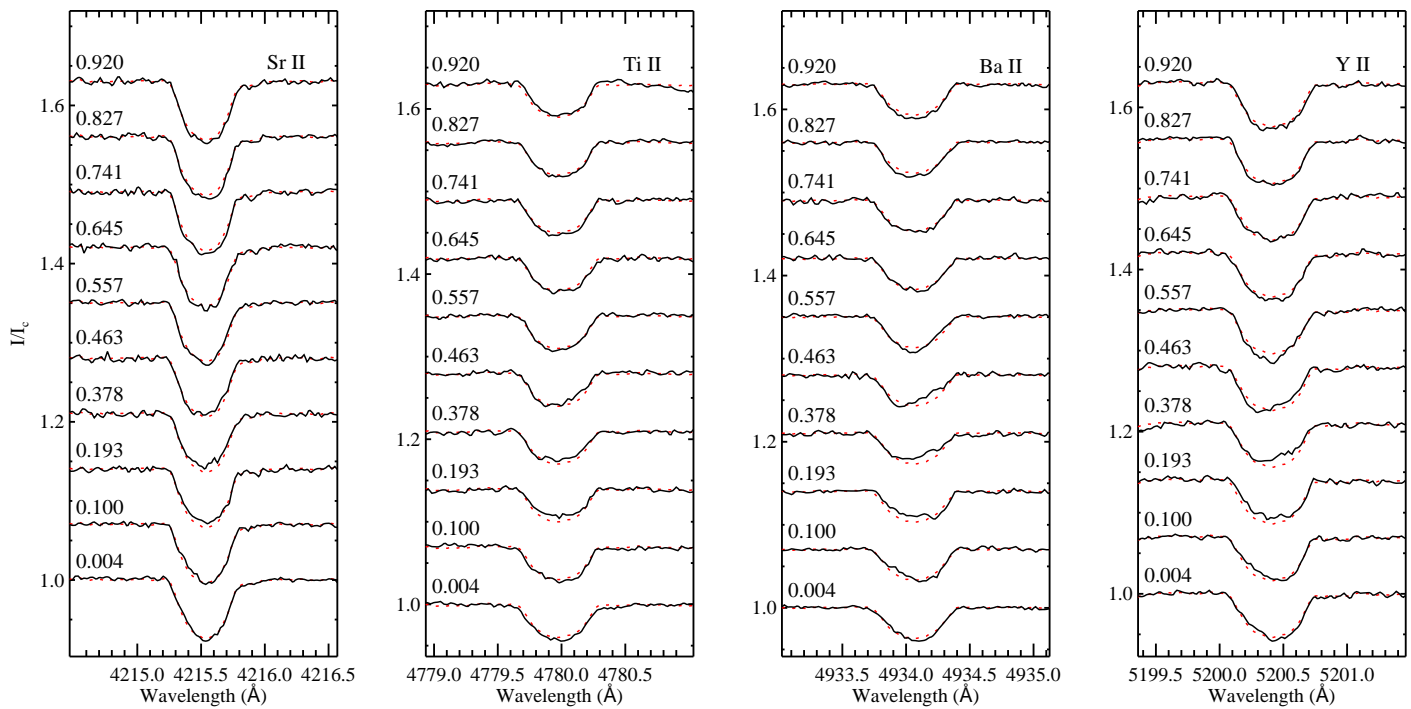}}}}
  \caption{Illustration of the line profile variability in the primary component (HgMn star) of 66~Eri. On the panels from left to right we show spectral lines as follows: \ion{Sr}{ii}~4215.159~\AA, \ion{Ti}{ii}~4779.985~\AA, \ion{Ba}{ii}~4934.076~\AA, and \ion{Y}{ii}~5200.410~\AA. The dashed line represents an average profile. All profiles are shifted upwards for displaying purposes.}
  \label{varprofs}
\end{figure*}

In Fig.~\ref{varprofs} we show typical variable spectral lines of the four chemical elements with definite line profile changes. Spectra are plotted together with an average profile (dotted line). The strongest variability is found for Sr and Ba lines; the weakest is seen for Ti. Examination of the line profile variability showed that all spectral lines listed in the Table~\ref{tab:varprofs} show the largest profile distortion relative to the mean spectrum at phases 0.193 and 0.378. The variability of \ion{Ba}{ii} is also clearly seen for the phase 0.463. In addition, the lines of \ion{Y}{ii} and \ion{Sr}{ii} also marginally deviate from the corresponding average profiles at phases 0.827, 0.920 and 0.741, 0.827 respectively.

The observed spectral variability occurs with the period of the orbital motion. For all four chemical elements with variable lines we found a distinct change of the radial velocity and, especially, equivalent width with the orbital period. Since our observations cover only two rotation cycles, we cannot determine variability period with a high precision. However, we observed that the integral properties of variable lines and their shapes repeat closely after one orbital period. This strongly suggests that the orbital motion and the stellar rotation are synchronized.

\section{Doppler imaging}
 \label{DI}

To study surface distribution of chemical elements in 66~Eri A, we applied the Doppler imaging (DI) method. We used INVERS10 code \citep{Piskunov:2002} to determine abundance maps, ignoring magnetic field. This analysis was applied to several non-blended spectral lines of Ti and Y that exhibit a significant variability and to all available spectral lines of Sr and Ba. In total, we used two spectral lines of \ion{Sr}{ii}, two of \ion{Ba}{ii}, six lines of \ion{Y}{ii}, and nine spectral lines of \ion{Ti}{ii}. These lines are marked with bold font in Table~\ref{tab:varprofs}. Multi-line DI is applied here to study spots in an HgMn star for the first time. This allowed us to reconstruct the surface maps with a higher precision than what was usually achieved using a single spectral line.

\begin{figure*}[t]
   \centering
  {\resizebox{\hsize}{!}{\rotatebox{0}{\includegraphics{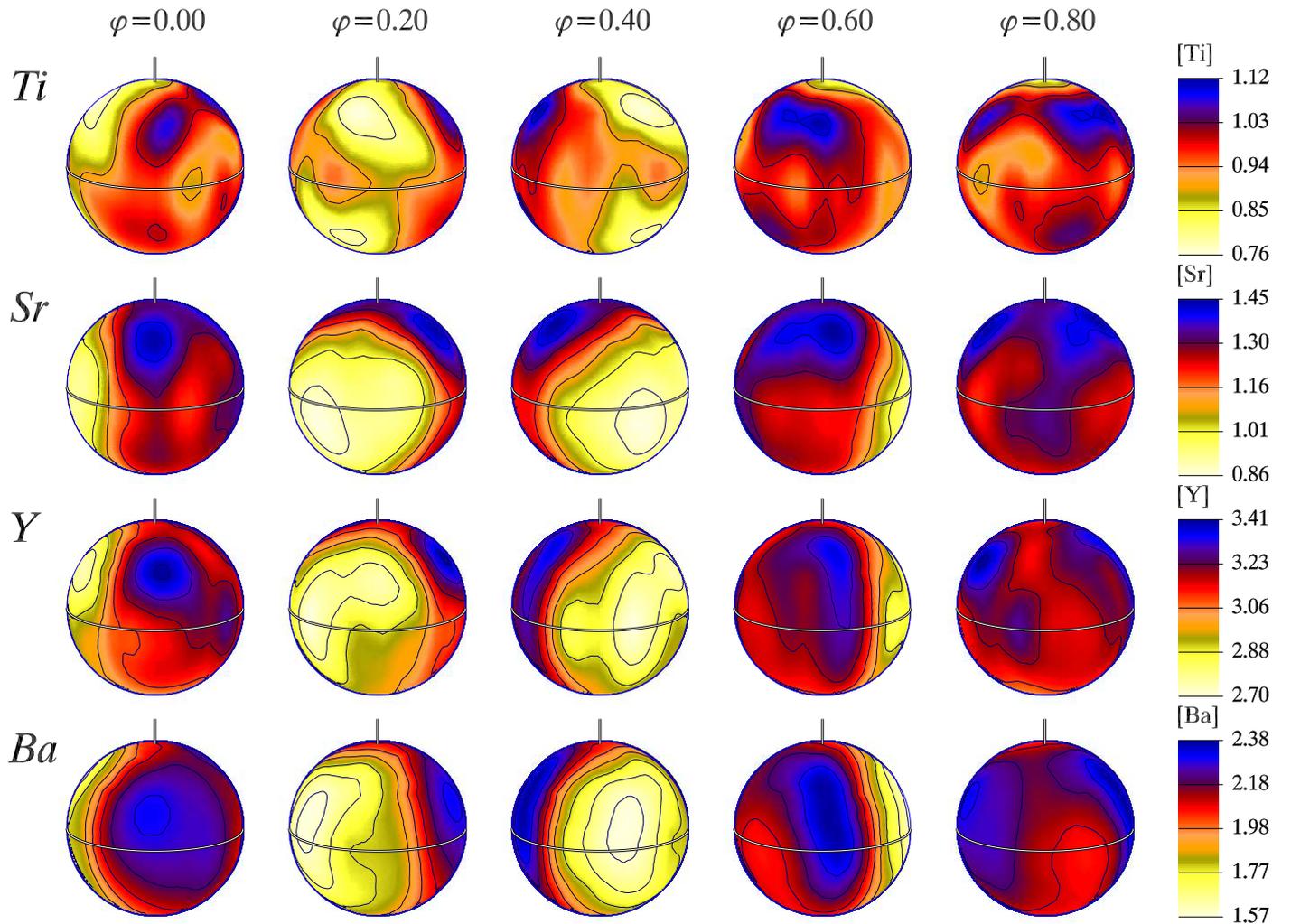}}}}
   \caption{Surface distributions of Ti, Sr, Y, and Ba abundances reconstructed for 66~Eri A. The star is shown at five equidistant rotation phases at the inclination angle $i=75$\degr\ and vertically oriented axis of rotation. The darker spots represent a higher concentrations of chemical elements. The contours of equal abundance are plotted with a step of 0.1 dex for all chemical species. The thick line indicates the position of rotational equator, while the vertical bar shows rotational pole. The scale of abundance maps is shown on the right. Abundances are given in a logarithmic scale relative to the Sun.}
   \label{DImaps}
\end{figure*}

Doppler inversions were performed using the inclination angle of 75\degr, as determined earlier (Sect.~\ref{fp}). The initial abundances of Ti, Sr, Y, and Ba were taken from \citet{Yuschenko:1999}. Spectra were calculated employing the revised model atmosphere of 66~Eri A with parameters \Teff$\,=11100$~K and $\log g=4.25$. The rotational velocity was adjusted to give a better fit to the observed line profiles, yielding the final value of \vsini$\,=17.5\pm0.2$~\kms. The reconstructed surface maps are shown in Fig.~\ref{DImaps}. Online Figs.~\ref{DI_fit1}--\ref{DI_fit3} illustrate the comparison of observed and model line profiles.

First of all, we can see from these maps that there is a general asymmetry of chemical composition for two hemispheres of the star. For all four studied chemical elements we find a larger abundance at rotational phases from 0.5 to 0.1 and a smaller abundance for phases 0.1--0.5. Secondly, there is a different spot structure for the hemisphere with element overabundances.

The spots of Ba and other chemical elements except Ti extend from $-30$\degr\ to 60\degr\ in latitude, showing the highest concentrations at the phase 0.65. For Y and Sr we also find the secondary spot at the phases 0.95--1.05. The spots of Sr and Ti are more extended azimuthally, while the spots of Ba and Y are stretched meridionally. Sr is also the only element of the four that shows a relative overabundance at the pole. Finally, Ti shows the spot positions similar to Sr, but with a discontinuity at the stellar equator even for the region of relative underabundance.

For all chemical elements presented in Fig.~\ref{DImaps}, the abundance difference between zones of the highest and lowest concentrations does not exceed 1~dex. For Ba the range is 0.81~dex, for Y it is 0.71~dex. For Ti and Sr the range is 0.36~dex and 0.59~dex, respectively. This is in a good agreement with the relative amplitude of variability in individual line profiles of these chemical elements (Fig.~\ref{varprofs}).

To exclude the possibility that spots are found by DI code at roughly one hemisphere of the star due to a spurious systematic effect, we applied the same DI analysis to ten spectral lines of Fe, which show no variability. The inferred abundance map of Fe is not similar to the maps of other chemical elements in its morphology. In particular, it does not show an abundance difference between two hemispheres. The difference between the regions of the highest and lowest abundances does not exceed $\approx0.15$~dex, which we take to be an upper limit of the possible inhomogeneity for chemical elements without obvious line profile variability.

\onlfig{8}{
\begin{figure*}
\centering
 \includegraphics[width=8cm]{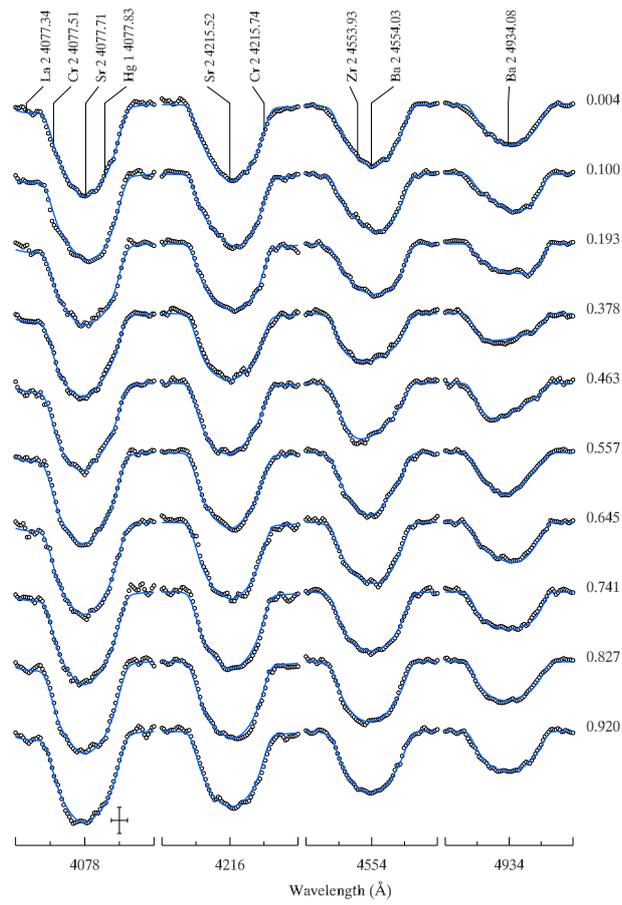}
 \caption{Comparison of the observed ({\it symbols}) and model ({\it solid lines}) profiles for Sr and Ba lines in 66~Eri~A. The bars in the lower left corner indicate the horizontal and vertical scales (0.1~\AA\ and 5\% of the continuum intensity respectively).}
 \label{DI_fit1}
\end{figure*}}

\onlfig{9}{
\begin{figure*}
\centering
 \includegraphics[width=10cm]{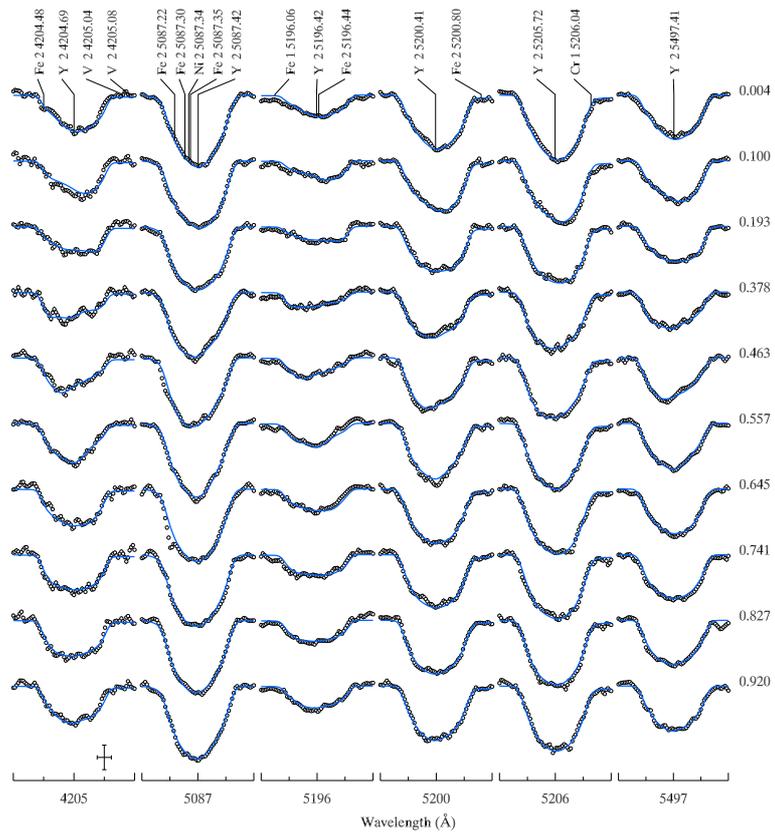}
 \caption{Same as Fig.~\ref{DI_fit1} but for Y.}
 \label{DI_fit2}
\end{figure*}}

\onlfig{10}{
\begin{figure*}
\centering
 \includegraphics[width=10cm]{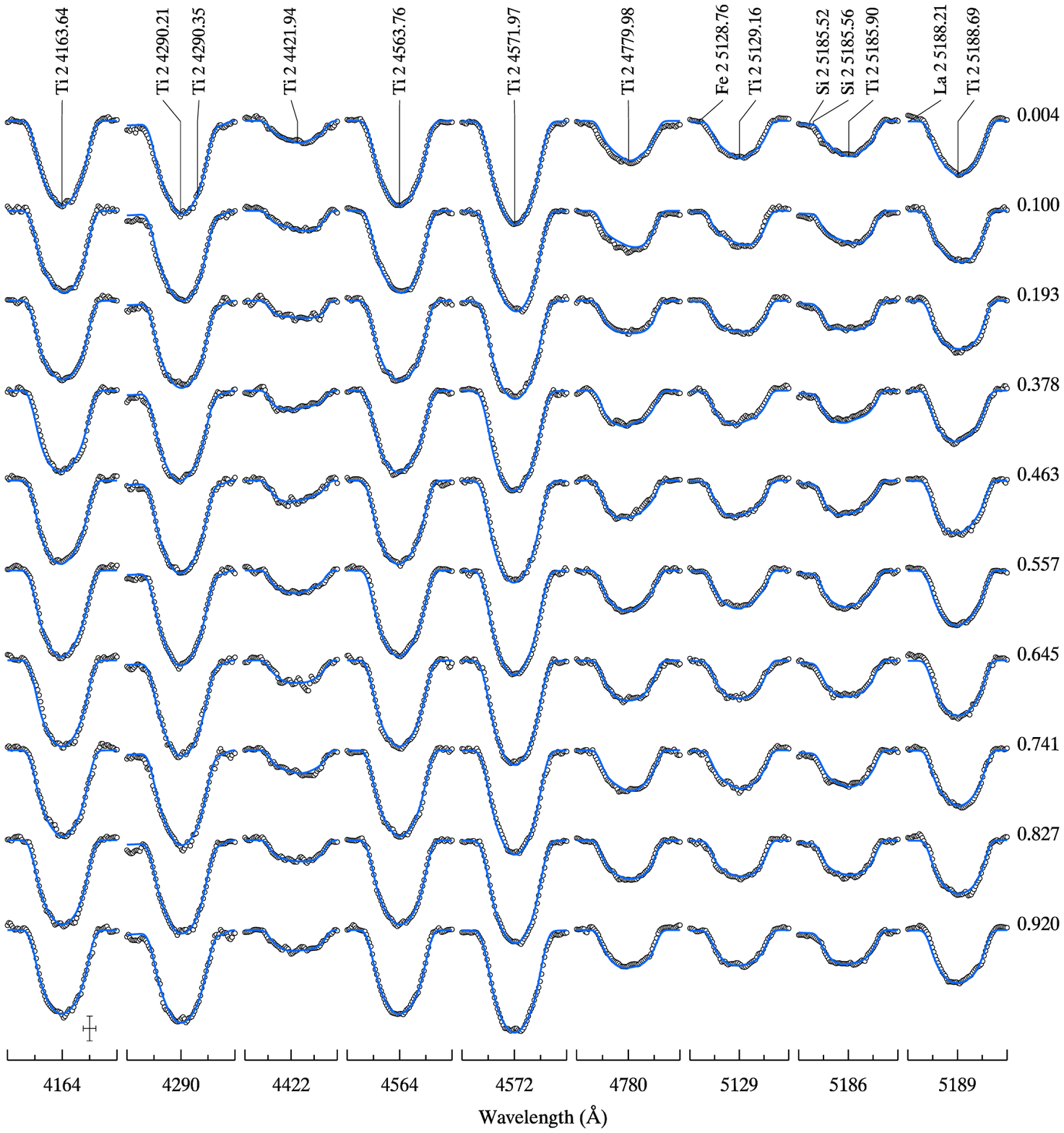}
 \caption{Same as Fig.~\ref{DI_fit1}  but for Ti.}
 \label{DI_fit3}
\end{figure*}}

\section{Summary and discussion}
 \label{disc}

Earlier studies of several HgMn stars found a variability in some spectral lines, attributing it to the presence of chemical spots. However, all previous attempts to find magnetic fields that can be responsible for these chemical inhomogeneities were unsuccessful \citep{Shorlin:2002, Wade:2006, Folsom:2010, Auriere:2010, Makaganiuk:2011}.

To extend the group of spotted HgMn stars with known constraints on magnetic field, we performed time-series spectropolarimetric observations of 66~Eri. This is the first such analysis of this unique spectroscopic binary star. It allowed us to search for magnetic fields in both components and investigate their line profile variability.

Taking the advantage of one of the best spectropolarimeters in the world, we recorded high $S/N$ and high-resolution spectra of 66~Eri covering its full orbital period. Despite the high polarimetric sensitivity of the instrument, we did not find any evidence of the longitudinal magnetic field in either component of 66~Eri. Our measurements of \Bz\ using LSD profiles have error bars 10--24~G for both stars. This null result agrees with the outcome of our earlier work \citep{Makaganiuk:2011}, where the longitudinal magnetic field was measured in a sample of 47 HgMn stars. The analysis of these objects showed no evidence of weak magnetic field signatures in Stokes~$V$ spectra.

Our magnetic field measurements provide an upper limit of 60--70~G for a dipolar field in either star and LSD~$V$ profiles provide no evidence for the presence of strong, complex magnetic field structures. This upper limit is significantly smaller than the minimum dipolar field diagnosed in Ap stars \citep{Auriere:2007}. Thus, 66~Eri is clearly not an Ap star.

As we deal with the binary system that shows two systems of spectral lines in its composite spectrum, it is necessary to disentangle the spectra to study each component separately. This was accomplished with the procedure of spectral disentangling. Along with the intrinsic spectra of the components it provided their radial velocities. The fit of our RV measurements together with RVs from previous studies \citep{Young:1976, Yuschenko:1999, Yuschenko:2001, Catanzaro:2004} with theoretical model, enabled us to improve the orbital elements of 66~Eri.

Using constraints from the orbital solution, we found the fundamental parameters for each star with the help of evolutionary tracks. The primary (HgMn) star turned to be more luminous and hotter, as expected. \citet{Yuschenko:1999} considered HgMn star to be the secondary component in their abundance analysis. The effective temperature revision of 200~K suggested in our study is relatively unimportant for abundance determination.

The evolutionary tracks provided masses of the components, which were utilized to compute the orbital inclination angle. The other method, based on the oblique rotator model, yielded nearly the same value for the rotational inclination angles. These results indicate that the angle between the rotation axes of two stars and the orbital plane is $90\degr$ and their rotation is synchronized with the orbital motion.

For the identification of spectral lines, we computed a synthetic spectrum based on the abundances provided by \citet{Yuschenko:1999} for both components. For the primary component the agreement between our synthetic spectrum and observations is rather good in most cases, in particular for the lines of Fe, Ti, Cr, Y, Ba, Si, Mn, and Hg. At the same time, we cannot confirm the high abundances of La, Zr, Yb, Hf, Pt, Au, and W inferred by \citet{Yuschenko:1999}. The strongest lines of these elements are very weak or absent in the observations while in synthetic spectrum they are too strong. The abundances of all these elements, derived by \citet{Yuschenko:1999} based on a small number of very weak spectral lines, are not reliable. The abundances for the secondary component seem to be in a very good agreement with our observations.

In conclusion of their abundance analysis \citet{Yuschenko:1999} made a confusing suggestion that the more massive component of 66~Eri is not an HgMn star. This claim contradicts the standard definition of an HgMn star as a late-B, slowly rotating object with enhanced Hg or Mn lines \citep[e.g.][]{Preston:1974}. \citet{Yuschenko:1999} themselves found both Hg and Mn to be overabundant; a large Hg overabundance was confirmed by \citet{Woolf:1999}. The absence of P overabundance is unusual, but is known for several other HgMn stars with \Teff\ similar to 66~Eri~A \citep{Adelman:2001a}, while a significant Ga overabundance is mostly inferred from the optical spectra of HgMn stars much hotter than 66~Eri~A \citep{Ryabchikova:1996}. Thus, 66~Eri~A is undoubtedly an HgMn star and was treated as such in many modern studies \citep{Cowley:2007, Catanzaro:2004, Rottler:2002, Woolf:1999, Hubrig:1998}. 66~Eri is also classified as ``Hg'' in the updated ``Catalogue of Ap, HgMn and Am stars'' \citep{Renson:2009}.

The analysis of spectral lines in the disentangled spectra of 66~Eri A allowed us to discover the variability in some of them. This was not achieved in the previous studies of this star because of a low resolving power and poor $S/N$. On the other hand, same analysis revealed no variable spectral lines for 66~Eri B.

Spectroscopic variability of 66~Eri A turned out to be different from other HgMn stars studied previously with time-series spectroscopy. All of them show notable variability in \ion{Hg}{ii} spectral lines. For example, for $\alpha$~And \citet{Kochukhov:2007} reported $\approx30\%$ equivalent width changes of the \ion{Hg}{ii} 3984~\AA\ line. The variability in this line is barely detectable in 66~Eri A. Our spectra are also insufficient to unambiguously show variability in Cr lines, although we found marginal changes in four of them.

More pronounced variability is found for the spectral lines of Y, Sr, Ba, and Ti. In Fig.~\ref{varprofs} we illustrate the line profiles of these four chemical elements that possess the strongest variability among all spectral lines of each species. The change in profiles with the rotational phase is still rather weak compared to other spotted HgMn stars, which makes this star the least variable.

We found slightly different line profile variability pattern for four elements. This indicates difference in the distribution of these chemical elements over the surface of 66~Eri A. We applied Doppler imaging technique to study spot geometry. The derived surface maps (Fig.~\ref{DImaps}) indicate higher concentrations of chemical elements on one hemisphere of the star. The overall abundance contrast is about 0.5 dex for all studied chemical elements. The hemisphere with a lower abundance still has higher concentrations of these elements compared to the Sun. This means that a certain physical process leads to a systematic difference in abundances between the two hemispheres. This phenomenon cannot be attributed to the strong magnetic field due to its absence in the atmosphere of 66~Eri A.

Low-contrast abundance inhomogeneities found on the surface of 66~Eri~A may give rise to photometric variability. So far the evidence of light variation in 66~Eri was controversial. \citet{Yuschenko:1999} could not verify weak photometric variability suggested by \citep{Schneider:1987}, but suggested the presence of variability with half of the orbital period in the Hipparcos epoch photometry of 66~Eri. Our reanalysis of these data does not support the claim by Yushchenko et al. There is a peak in the amplitude spectrum at $P=2.754\pm0.004$~d, which is close to, but formally is not consistent with, $P_{\rm orb}/2=2.761$~d. However, this periodicity is not statistically significant (FAP\,=\,0.4) and there are more than ten other peaks in the amplitude spectrum with the same or higher magnitude.

\begin{figure*}[t]
   \centering
  {\resizebox{8cm}{!}{\rotatebox{0}{\includegraphics{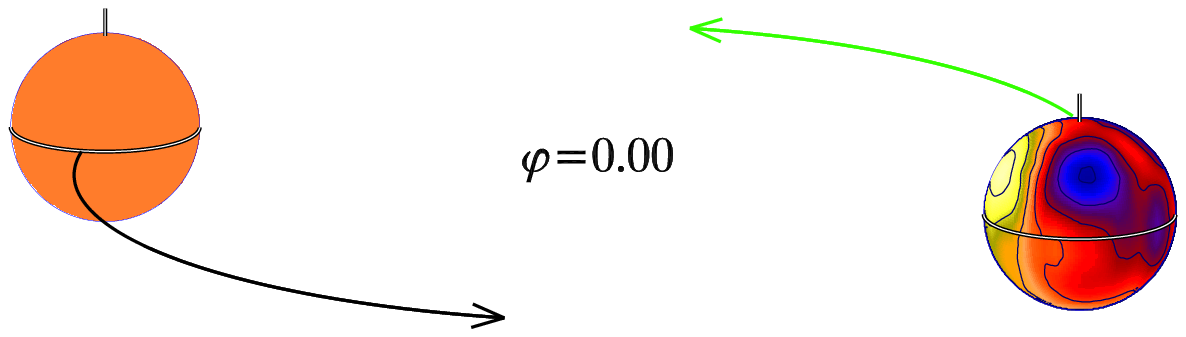}}}}
  {\resizebox{8cm}{!}{\rotatebox{0}{\includegraphics{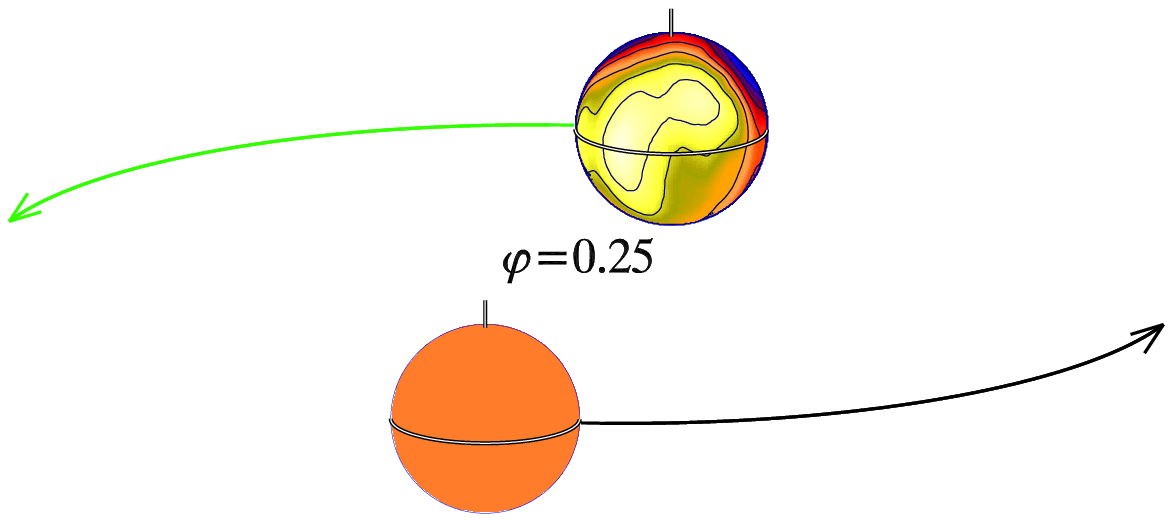}}}}
  {\resizebox{8cm}{!}{\rotatebox{0}{\includegraphics{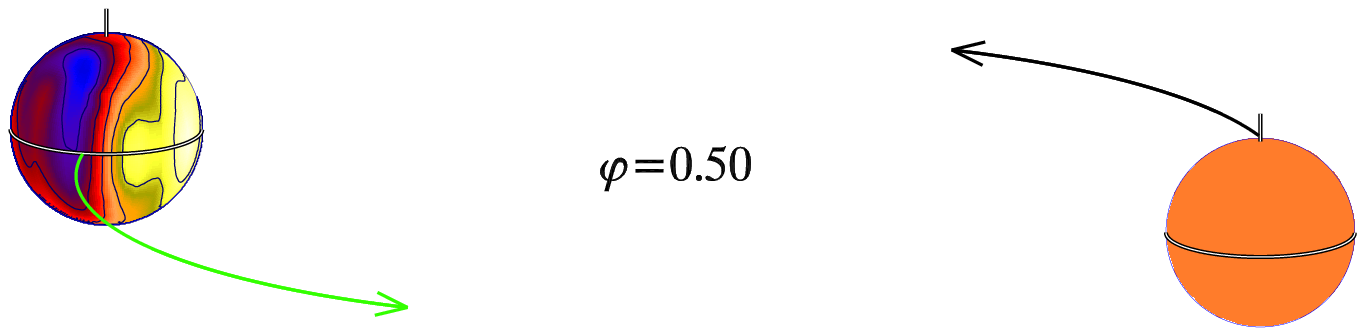}}}}
  {\resizebox{8cm}{!}{\rotatebox{0}{\includegraphics{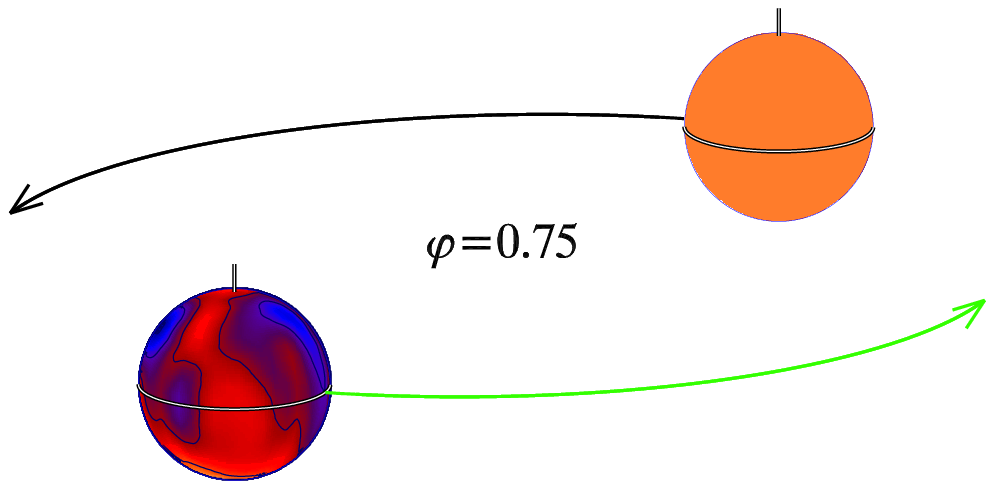}}}}
   \caption{Orbital motion and rotation of the components in the binary system 66~Eri. The star with yttrium spots represents 66~Eri A and the other filled with a solid color represents 66~Eri B. The radii and the separation between the two stars are shown to scale.}
   \label{DIrot}
\end{figure*}

Since 66~Eri is a binary system, we cannot exclude the influence of the secondary component on the atmosphere of the primary. To understand how the position of chemical spots in 66~Eri A correlates with the relative orbital motion of the components, we combined the results of the orbital analysis and DI maps of yttrium. The resulting picture is shown in Fig.~\ref{DIrot}.

It turns out that the relative underabundance of Ti, Sr, Y, and Ba in 66~Eri A is observed on the hemisphere which faces 66~Eri B. Due to a rather close distance between the two stars, the tidal effects can be responsible for the destruction of the upper layer containing the excess of studied elements. On the other hand, the radiation flux from 66~Eri B can also have an impact on the distribution of chemical elements in the atmosphere of the primary component. However, it is not clear why is only one of the nearly identical stars affected by these effects.

In addition to the puzzling spectroscopic variability in 66~Eri~A, the two stars have significantly different average chemical characteristics. Despite the two stars having nearly the same fundamental characteristics, the secondary component seems to be a normal star with nearly solar chemical composition and shows no surface inhomogeneities. Remarkably, the line profile variability in A component occurs only for those elements which show much stronger spectral lines in the primary. This is also clearly seen from the abundances of Ti, Sr, Y, and Ba determined by \citet{Yuschenko:1999}: all these four elements are significantly more abundant in the atmosphere of the primary. Among the elements with the largest abundance difference between the components only Hg does not clearly show a non-uniform distribution.

The identical age of both stars and the small separation between them suggest that they were formed in the same protostellar gas cloud. This would imply the same overall chemical composition for both stars. The observed chemical inhomogeneities in one of the two components and a large difference between their average abundance could be related. These effects cannot be attributed to the generic atomic diffusion, which depends only on fundamental stellar parameters. It must be a result of some rapid process that changes the atomic diffusion on a timescale much shorter than the age of a star. If this is true, the secondary component has either already undergone the stage of non-uniform surface distribution of chemical elements or may show surface structures in the future.

\bibliographystyle{aa}
\bibliography{references}

\begin{thebibliography}{45}
\expandafter\ifx\csname natexlab\endcsname\relax\def\natexlab#1{#1}\fi

\bibitem[{{Adelman} {et~al.}(2002){Adelman}, {Gulliver}, {Kochukhov}, \&
  {Ryabchikova}}]{Adelman:2002}
{Adelman}, S.~J., {Gulliver}, A.~F., {Kochukhov}, O.~P., \& {Ryabchikova},
  T.~A. 2002, \apj, 575, 449

\bibitem[{{Adelman} {et~al.}(2001){Adelman}, {Gulliver}, \&
  {Rayle}}]{Adelman:2001a}
{Adelman}, S.~J., {Gulliver}, A.~F., \& {Rayle}, K.~E. 2001, \aap, 367, 597

\bibitem[{{Auri{\`e}re} {et~al.}(2009){Auri{\`e}re}, {Wade},
  {Konstantinova-Antova}, {Charbonnel}, {Catala}, {Weiss}, {Roudier}, {Petit},
  {Donati}, {Alecian}, {Cabanac}, {van Eck}, {Folsom}, \&
  {Power}}]{Auriere:2009}
{Auri{\`e}re}, M., {Wade}, G.~A., {Konstantinova-Antova}, R., {et~al.} 2009,
  \aap, 504, 231

\bibitem[{{Auri{\`e}re} {et~al.}(2010){Auri{\`e}re}, {Wade}, {Ligni{\`e}res},
  {Hui-Bon-Hoa}, {Landstreet}, {Iliev}, {Donati}, {Petit}, {Roudier}, \&
  {Th{\'e}ado}}]{Auriere:2010}
{Auri{\`e}re}, M., {Wade}, G.~A., {Ligni{\`e}res}, F., {et~al.} 2010, \aap,
  523, A40

\bibitem[{{Auri{\`e}re} {et~al.}(2007){Auri{\`e}re}, {Wade}, {Silvester},
  {Ligni{\`e}res}, {Bagnulo}, {Bale}, {Dintrans}, {Donati}, {Folsom},
  {Gruberbauer}, {Hui Bon Hoa}, {Jeffers}, {Johnson}, {Landstreet},
  {L{\`e}bre}, {Lueftinger}, {Marsden}, {Mouillet}, {Naseri}, {Paletou},
  {Petit}, {Power}, {Rincon}, {Strasser}, \& {Toqu{\'e}}}]{Auriere:2007}
{Auri{\`e}re}, M., {Wade}, G.~A., {Silvester}, J., {et~al.} 2007, \aap, 475,
  1053

\bibitem[{{Bagnulo} {et~al.}(2009){Bagnulo}, {Landolfi}, {Landstreet}, {Landi
  Degl'Innocenti}, {Fossati}, \& {Sterzik}}]{Bagnulo:2009}
{Bagnulo}, S., {Landolfi}, M., {Landstreet}, J.~D., {et~al.} 2009, \pasp, 121,
  993

\bibitem[{{Berghofer} \& {Schmitt}(1994)}]{Berghofer:1994}
{Berghofer}, T.~W. \& {Schmitt}, J.~H.~M.~M. 1994, \aap, 292, L5

\bibitem[{{Briquet} {et~al.}(2010){Briquet}, {Korhonen}, {Gonz{\'a}lez},
  {Hubrig}, \& {Hackman}}]{Briquet:2010}
{Briquet}, M., {Korhonen}, H., {Gonz{\'a}lez}, J.~F., {Hubrig}, S., \&
  {Hackman}, T. 2010, \aap, 511, A71

\bibitem[{{Catanzaro} \& {Leto}(2004)}]{Catanzaro:2004}
{Catanzaro}, G. \& {Leto}, P. 2004, \aap, 416, 661

\bibitem[{{Cowley} {et~al.}(2007){Cowley}, {Hubrig}, {Castelli},
  {Gonz{\'a}lez}, \& {Wolff}}]{Cowley:2007}
{Cowley}, C.~R., {Hubrig}, S., {Castelli}, F., {Gonz{\'a}lez}, J.~F., \&
  {Wolff}, B. 2007, \mnras, 377, 1579

\bibitem[{{Donati} \& {Landstreet}(2009)}]{Donati:2009}
{Donati}, J. \& {Landstreet}, J.~D. 2009, \araa, 47, 333

\bibitem[{{Donati} {et~al.}(1997){Donati}, {Semel}, {Carter}, {Rees}, \&
  {Collier Cameron}}]{Donati:1997}
{Donati}, J., {Semel}, M., {Carter}, B.~D., {Rees}, D.~E., \& {Collier
  Cameron}, A. 1997, \mnras, 291, 658

\bibitem[{{Folsom} {et~al.}(2010){Folsom}, {Kochukhov}, {Wade}, {Silvester}, \&
  {Bagnulo}}]{Folsom:2010}
{Folsom}, C.~P., {Kochukhov}, O., {Wade}, G.~A., {Silvester}, J., \& {Bagnulo},
  S. 2010, \mnras, 407, 2383

\bibitem[{{Frost} \& {Struve}(1924)}]{Frost:1924}
{Frost}, E.~B. \& {Struve}, O. 1924, \apj, 60, 313

\bibitem[{{Girardi} {et~al.}(2000){Girardi}, {Bressan}, {Bertelli}, \&
  {Chiosi}}]{Girardi:2000}
{Girardi}, L., {Bressan}, A., {Bertelli}, G., \& {Chiosi}, C. 2000, \aaps, 141,
  371

\bibitem[{{Hubrig}(1998)}]{Hubrig:1998}
{Hubrig}, S. 1998, Contributions of the Astronomical Observatory Skalnate
  Pleso, 27, 296

\bibitem[{{Hubrig} {et~al.}(2006){Hubrig}, {Gonz{\'a}lez}, {Savanov},
  {Sch{\"o}ller}, {Ageorges}, {Cowley}, \& {Wolff}}]{Hubrig:2006}
{Hubrig}, S., {Gonz{\'a}lez}, J.~F., {Savanov}, I., {et~al.} 2006, \mnras, 371,
  1953

\bibitem[{{Hubrig} {et~al.}(2001){Hubrig}, {Le Mignant}, {North}, \&
  {Krautter}}]{Hubrig:2001}
{Hubrig}, S., {Le Mignant}, D., {North}, P., \& {Krautter}, J. 2001, \aap, 372,
  152

\bibitem[{{Khokhlova} {et~al.}(1995){Khokhlova}, {Zverko}, {Zhizhnovskii}, \&
  {Griffin}}]{Hohlova:1995}
{Khokhlova}, V.~L., {Zverko}, Y., {Zhizhnovskii}, I., \& {Griffin}, R.~E.~M.
  1995, Astronomy Letters, 21, 818

\bibitem[{{Kochukhov} {et~al.}(2007){Kochukhov}, {Adelman}, {Gulliver}, \&
  {Piskunov}}]{Kochukhov:2007}
{Kochukhov}, O., {Adelman}, S.~J., {Gulliver}, A.~F., \& {Piskunov}, N. 2007,
  Nature Physics, 3, 526

\bibitem[{{Kochukhov} {et~al.}(2010){Kochukhov}, {Makaganiuk}, \&
  {Piskunov}}]{Kochukhov:2010}
{Kochukhov}, O., {Makaganiuk}, V., \& {Piskunov}, N. 2010, \aap, 524, A5

\bibitem[{{Kochukhov} {et~al.}(2005){Kochukhov}, {Piskunov}, {Sachkov}, \&
  {Kudryavtsev}}]{Kochukhov:2005}
{Kochukhov}, O., {Piskunov}, N., {Sachkov}, M., \& {Kudryavtsev}, D. 2005,
  \aap, 439, 1093

\bibitem[{{Leroy} {et~al.}(1994){Leroy}, {Bagnulo}, {Landolfi}, \& {Landi
  Degl'Innocenti}}]{Leroy:1994}
{Leroy}, J.~L., {Bagnulo}, S., {Landolfi}, M., \& {Landi Degl'Innocenti}, E.
  1994, \aap, 284, 174

\bibitem[{{Ligni{\`e}res} {et~al.}(2009){Ligni{\`e}res}, {Petit}, {B{\"o}hm},
  \& {Auri{\`e}re}}]{Lignieres:2009}
{Ligni{\`e}res}, F., {Petit}, P., {B{\"o}hm}, T., \& {Auri{\`e}re}, M. 2009,
  \aap, 500, L41

\bibitem[{{Makaganiuk} {et~al.}(2011){Makaganiuk}, {Kochukhov}, {Piskunov},
  {Jeffers}, {Johns-Krull}, {Keller}, {Rodenhuis}, {Snik}, {Stempels}, \&
  {Valenti}}]{Makaganiuk:2011}
{Makaganiuk}, V., {Kochukhov}, O., {Piskunov}, N., {et~al.} 2011, \aap, 525,
  A97

\bibitem[{{Mayor} {et~al.}(2003){Mayor}, {Pepe}, {Queloz}, {Bouchy},
  {Rupprecht}, {Lo Curto}, {Avila}, {Benz}, {Bertaux}, {Bonfils}, {Dall},
  {Dekker}, {Delabre}, {Eckert}, {Fleury}, {Gilliotte}, {Gojak}, {Guzman},
  {Kohler}, {Lizon}, {Longinotti}, {Lovis}, {Megevand}, {Pasquini}, {Reyes},
  {Sivan}, {Sosnowska}, {Soto}, {Udry}, {van Kesteren}, {Weber}, \&
  {Weilenmann}}]{HARPS}
{Mayor}, M., {Pepe}, F., {Queloz}, D., {et~al.} 2003, The Messenger, 114, 20

\bibitem[{{Piskunov} \& {Kochukhov}(2002)}]{Piskunov:2002}
{Piskunov}, N. \& {Kochukhov}, O. 2002, \aap, 381, 736

\bibitem[{{Piskunov} {et~al.}(1995){Piskunov}, {Kupka}, {Ryabchikova}, {Weiss},
  \& {Jeffery}}]{VALD}
{Piskunov}, N.~E., {Kupka}, F., {Ryabchikova}, T.~A., {Weiss}, W.~W., \&
  {Jeffery}, C.~S. 1995, \aaps, 112, 525

\bibitem[{{Piskunov} \& {Valenti}(2002)}]{reduce}
{Piskunov}, N.~E. \& {Valenti}, J.~A. 2002, \aap, 385, 1095

\bibitem[{{Preston}(1974)}]{Preston:1974}
{Preston}, G.~W. 1974, \araa, 12, 257

\bibitem[{{Renson} \& {Manfroid}(2009)}]{Renson:2009}
{Renson}, P. \& {Manfroid}, J. 2009, \aap, 498, 961

\bibitem[{{Rottler} \& {Young}(2002)}]{Rottler:2002}
{Rottler}, L. \& {Young}, A. 2002, in Bulletin of the American Astronomical
  Society, Vol. 201, Bulletin of the American Astronomical Society, 404

\bibitem[{{Ryabchikova} {et~al.}(1996){Ryabchikova}, {Zakharova}, \&
  {Adelman}}]{Ryabchikova:1996}
{Ryabchikova}, T.~A., {Zakharova}, L.~A., \& {Adelman}, S.~J. 1996, \mnras,
  283, 1115

\bibitem[{{Schneider}(1987)}]{Schneider:1987}
{Schneider}, H. 1987, Hvar Observatory Bulletin, 11, 29

\bibitem[{{Shorlin} {et~al.}(2002){Shorlin}, {Wade}, {Donati}, {Landstreet},
  {Petit}, {Sigut}, \& {Strasser}}]{Shorlin:2002}
{Shorlin}, S.~L.~S., {Wade}, G.~A., {Donati}, J., {et~al.} 2002, \aap, 392, 637

\bibitem[{{Snik} {et~al.}(2008){Snik}, {Jeffers}, {Keller}, {Piskunov},
  {Kochukhov}, {Valenti}, \& {Johns-Krull}}]{HARPSpol}
{Snik}, F., {Jeffers}, S., {Keller}, C., {et~al.} 2008, in Society of
  Photo-Optical Instrumentation Engineers (SPIE) Conference Series, Vol. 7014,
  22

\bibitem[{{Snik} {et~al.}(2010){Snik}, {Kochukhov}, {Piskunov}, {Rodenhuis},
  {Jeffers}, {Keller}, {Dolgopolov}, {Stempels}, {Makaganiuk}, {Valenti}, \&
  {Johns-Krull}}]{Snik:2010}
{Snik}, F., {Kochukhov}, O., {Piskunov}, N., {et~al.} 2010, ArXiv e-prints

\bibitem[{{Stelzer} {et~al.}(2006){Stelzer}, {Hu{\'e}lamo}, {Micela}, \&
  {Hubrig}}]{Stelzer:2006}
{Stelzer}, B., {Hu{\'e}lamo}, N., {Micela}, G., \& {Hubrig}, S. 2006, \aap,
  452, 1001

\bibitem[{{van Leeuwen} \& {Fantino}(2005)}]{Van-Leeuwen:2005}
{van Leeuwen}, F. \& {Fantino}, E. 2005, \aap, 439, 791

\bibitem[{{Wade} {et~al.}(2006){Wade}, {Auri{\`e}re}, {Bagnulo}, {Donati},
  {Johnson}, {Landstreet}, {Ligni{\`e}res}, {Marsden}, {Monin}, {Mouillet},
  {Paletou}, {Petit}, {Toqu{\'e}}, {Alecian}, \& {Folsom}}]{Wade:2006}
{Wade}, G.~A., {Auri{\`e}re}, M., {Bagnulo}, S., {et~al.} 2006, \aap, 451, 293

\bibitem[{{Wade} {et~al.}(2000){Wade}, {Donati}, {Landstreet}, \&
  {Shorlin}}]{Wade:2000}
{Wade}, G.~A., {Donati}, J., {Landstreet}, J.~D., \& {Shorlin}, S.~L.~S. 2000,
  \mnras, 313, 823

\bibitem[{{Woolf} \& {Lambert}(1999)}]{Woolf:1999}
{Woolf}, V.~M. \& {Lambert}, D.~L. 1999, \apj, 521, 414

\bibitem[{{Young}(1976)}]{Young:1976}
{Young}, A. 1976, \pasp, 88, 275

\bibitem[{{Yushchenko} {et~al.}(1999){Yushchenko}, {Gopka}, {Khokhlova},
  {Musaev}, \& {Bikmaev}}]{Yuschenko:1999}
{Yushchenko}, A.~V., {Gopka}, V.~F., {Khokhlova}, V.~L., {Musaev}, F.~A., \&
  {Bikmaev}, I.~F. 1999, Astronomy Letters, 25, 453

\bibitem[{{Yushchenko} {et~al.}(2001){Yushchenko}, {Gopka}, {Khokhlova}, \&
  {Tomkin}}]{Yuschenko:2001}
{Yushchenko}, A.~V., {Gopka}, V.~F., {Khokhlova}, V.~L., \& {Tomkin}, J. 2001,
  Information Bulletin on Variable Stars, 5213, 1

\end{thebibliography}

\end{document}